# Mapping the Depths: A Stocktake of Underground Power Distribution in United States

Tao Sun[*], Chad Zanocco[*], June Flora[*], Ram Rajagopal[*†]
[*]Department of Civil and Environmental Engineering, Stanford University, CA, USA
[†]Department of Electrical Engineering, Stanford University, CA, USA

*Abstract*—A resilient energy infrastructure is crucial for addressing increasing extreme weather and climate risks. The undergrounding of the power system is one approach to building such resiliency. In this study, we introduce Grid Underground Distribution Statistics (GUDS) for the US, the first nationwide comprehensive assessment of underground electricity distribution at a high spatial granularity. In analyzing this dataset, we find regional differences in underground distribution rates, with generally higher rates for east and west coasts and in northern states, and lower rates in the central US. We also observe relationships between underground rates and factors such as household income levels, degree of urbanization, and vulnerability to natural hazards. Notably, regions with higher electricity rates are not associated with greater proportions of underground distribution, highlighting potential equity issues in infrastructure distribution. By presenting this granular information and insights on underground distribution, our study offers valuable guidance for informing planning and decision-making by policymakers, Independent System Operators, utilities, and end-users.

*Index Terms*--underground distribution, energy infrastructure, power resilience, climate change, extreme weather

## I. Introduction

The increasing frequency and intensity of extreme weather events, exacerbated by climate change, pose substantial threats to the reliability of energy systems [1]. Over the past two decades, there has been a marked rise in major power outages attributable to such extreme weather events [2]. One widely advocated solution is the undergrounding of new and existing above-ground power lines [3]. These buried lines, shielded from extreme weather events like hurricanes, blizzards, thunderstorms, and wildfires, offer a significant reduction in outage risks, offering a safer and more consistent distribution of electricity to end users.

Surprisingly, despite its merits, only about 18% of the U.S. distribution system is currently undergrounded, and the often-cited reasoning is due to the prohibitive costs of such plans. Undergrounding can represent a significant expenditure with substantially higher costs (up to five to ten times more per mile) compared to overhead installations [4]. Given both the higher expected benefits in terms of increased grid resilience and fewer outages, as well as high costs in terms of installation, it is crucial that undergrounding is strategically deployed to ensure economic efficiency and equity while focusing on the most vulnerable areas.

Recognizing this pressing need, U.S. utilities have initiated plans to increase the extent of their underground distribution systems. For instance, Pacific Gas & Electric Co. (PG&E) intends to transition 10,000 mi of their above-ground lines in wildfire-prone zones to an underground setup [5]. Florida Power & Light Company (FPL), spurred to action by the devastating hurricane seasons of 2004-05, plans to earmark $1 billion annually from 2025 for overhead-to-underground conversions [6]. Wisconsin Public Service (WPS) has already achieved significant progress, transitioning 2000 miles of lines over the past eight years [7].

This evolving landscape highlights the imperative for a comprehensive stocktake of the current undergrounding status across the U.S. to enable utilities, regulators, and consumers to make data-informed decisions. This overview provides critical insights into best practices, policy implications, and community resilience. However, the existing literature remains limited to mostly local scenarios [8]–[10] and there is a notable lack of nationwide analyses, especially those examining undergrounding trends related to socioeconomic and environmental contexts of specific regions.

This study aims to fill this knowledge gap by making the following contributions:

1. We introduce Grid Underground Distribution Statistics (GUDS) for the US, a comprehensive, first-of-its-kind accounting of underground distribution across the nation, both at the granular (Zip Code Tabulation Area, or ZCTA) and aggregated (state) levels.
2. We investigate the interplay between undergrounding and various socio-economic and environmental metrics at the local level.
3. We identify potential inequalities between the uptake of underground distribution systems and the electricity rates consumers face.

[1]Emails: luke18@stanford.edu (T.Sun), ramr@stanford.edu (R.Rajagopal). This work was supported in part by the National Science Foundation through a CAREER award (#1554178) and by a Stanford Precourt Pioneering Project award.

In the following sections we first describe our methodology and data, then present our findings, and finally draw conclusions based on the data.

## II. Methods

The methodology encompasses the approaches and analytical techniques we employed to gather, process, and evaluate the GUDS dataset on underground distribution line information. The subsequent subsections describe in detail the processes involved in gathering underground distribution line data, relating utility information to geographical domains, relevant socio-demographic, electric rate, and National Risk Index data, and finally, the statistical tools used for analyzing data and generating important insights.

### A. Underground Distribution Line Data and Definitions

We collected a unique and extensive dataset capturing both overhead and underground distribution line distances across 2680 electric utilities spanning the contiguous US, Alaska, and Hawaii. This data spans a period of 29 years (1993 - 2021) and was collected using a range of diverse sources and methods including: annual Form 10-K reports submitted to the U.S. Securities and Exchange Commission, the U.S. Department of Agriculture Rural Utilities Service Form7, the UDI Directory of Electric Power Producers & Distributors [11], official online portals of utilities and state governments, direct inquiries to state public utility commissions, and outreach to individual utilities.

To provide context to the scale and resolution of our dataset, we benchmarked its coverage against data from the US Energy Information Administration (EIA) Form EIA-861. As of 2020, a breakdown of 1228 utilities—comprising 149 investor-owned utilities, 551 cooperatives, and 528 public power utilities—submitted their retail electricity sales on EIA-861. Our dataset covered a vast majority of these: 97% of investor-owned utilities, 99% of cooperatives, and 95% of public power utilities. The remaining utilities within our set of 2680 either had not reported in EIA-861 in 2020 or had ceased operations by 2020.

For the purposes of our research, we introduced the metric of Underground Distribution Rate (UDR) for a utility. This is calculated by taking the mileage of underground lines and dividing it by the combined mileage of both underground and overhead lines. Consequently, UDR values range between 0 and 1. While we have UDR values spanning multiple years based on our collected data, this study narrows its focus to the 2020 data (or the closest year available) to ensure that the most recent insights are highlighted.

### B. Utility to Geography: Underground Distribution at ZCTA and State Levels

We set out to align our utility-specific underground distribution data with different geographical territories, specifically the ZCTA and state levels. These frequently used geographical divisions allow us to combine our findings with existing social demographics and environmental metrics. Additionally, by downscaling to the ZCTA level, we can derive insights and trends that are more localized within larger territories. Since the service territories of utilities in our data cover all states and nearly all ZCTAs (more than 33,000) in the U.S., we are able to generate comprehensive geographical coverage of underground distribution data at these two levels.

To do so, we first obtained the ZCTA service regions corresponding to each utility in our dataset. This information was procured from NREL's "U.S. Electric Utility Companies and Rates: Look-up by Zip Code" dataset [12] and EIA-861's service territory records.

For each utility, we estimated the length of both underground and total distribution lines present within every ZCTA under its service. The allocation method is straightforward: the mileage of lines associated with a utility was distributed across its serving ZCTAs in proportion to the ZCTA's population. This approach relies on two primary assumptions: first, within a localized region, the length of distribution lines in an area is likely aligned with its customer count. Second, a utility's customer distribution across various ZCTAs is proportional to the ZCTA's population. These assumptions may not always hold, potentially causing some variation between the estimated and actual distribution data at the ZCTA level.

Subsequently, by aggregating the mileage data from all utilities serving a specific ZCTA, we could estimate its underground and total distribution line lengths. This allowed us to compute the UDR for each ZCTA. For a broader perspective, ZCTA-level data was further grouped by state, enabling us to estimate the UDR for each individual state.

### C. Social Demographics, Electric Rate, and NRI Data

We identified the following social demographics information at the ZCTA level: household mean income, population density, and Rural Urban Commuting Area (RUCA) Codes, a measure that distinguishes regions based on their urbanization and proximity to metro areas. Data sources for these measures included the US census and 2010 USDA-ERS Rural Urban Commuting Area Codes. A modification of these codes for finer classification, as described in [13], was adopted. The population density is calculated by dividing the total population by the land area (in square miles) for each ZCTA.

Regarding electricity cost metrics, we obtained the average residential rate at the ZCTA level. This data was also extracted from the NREL's dataset [12], titled "U.S. Electric Utility Companies and Rates: Look-up by Zipcode."

To assess community vulnerability to natural hazards, we incorporated the National Risk Index (NRI) as defined by the Federal Emergency Management Agency (FEMA) [14]. The NRI offers a holistic picture of U.S. communities' risk across 18 natural hazards. Initially available at the county level, the NRI data was subsequently mapped to individual ZCTAs based on their respective county affiliations.

### D. Statistical analysis

In order to investigate the relationships between the UDR and the selected sociodemographic and environmental factors, we conducted an Ordinary Least Squares (OLS) regression analysis.

The specific form of our OLS model is as follows:

$$UDP_i = \beta_0 + \beta_1 \times HouseholdIncome_i$$
$$+ \beta_2 \times PopulationDensity_i + \beta_3 \times RUCA_i$$
$$+ \beta_4 ElectricRate_i + \beta_5 \times NRI_i + \epsilon_i \quad (1)$$

Here $UDP_i$ represents the Underground Distribution Percent (UDP) for the $i^{th}$ observation or ZCTA (UDP equal to UDR multiplied by 100), $HouseholdIncome_i$, $PopulationDensity_i$, $RUCA_i$, $ElectricRate_i$, and $NRI_i$ are the predictors corresponding to household mean income, population density, rural urban commuting area code, electric rate, and NRI (value) respectively for the $i^{th}$ observation, $\beta_0$ is the intercept, $\beta_1$, $\beta_2$, … represent the coefficients of the predictors, and $\varepsilon_i$ is the error term.

In fitting the regression model in (1), we applied a log(1+x) transformation to both the dependent and independent variables. Consequently, the coefficients can be interpreted as the percentage change in the dependent variable corresponding to a 1% increase in the independent variable. The transformation was particularly beneficial for accommodating values close to zero.

### III. Results

In this section, we present the findings derived from our dataset and analysis detailed earlier in Section II. We explore the patterns of underground distribution lines, describe their geographical variation, and identify the relationship between factors that influence undergrounding.

#### A. Underground Distribution at ZCTA Level

In assessing the distribution infrastructure across the United States, it is crucial to explore the granular details at the ZCTA level, which allows for a more nuanced understanding of disparities, concentrations, and patterns.

Fig. 1 illustrates the diverse landscape of UDRs across the US at the ZCTA level. The Pacific West, including states like Washington, Oregon, and California, is mainly characterized by shades of cyan and green, indicating moderate UDRs, with some occasional blue patches near large municipalities. The Central US, with states like Nebraska, Kansas, and Oklahoma, is dominated by shades of yellow, denoting low distribution rates. The Northeast, including states like New York and Massachusetts, predominantly displays a gradient of cyans transitioning to blue, indicating areas of moderate to high underground distribution. Stretching along the Eastern coast from Virginia to Florida, there's a mix of yellow, cyan, green, and sporadic blue patches, reflecting a range of distribution rates. This higher UDR along the eastern coast aligns with typical hurricane paths in the area. Northern states such as North Dakota, Minnesota, and Wisconsin also exhibit a patchwork of cyan and green. This palette suggests a predominantly moderate rate of underground distribution, with certain areas leaning towards higher rates, potentially as a protective measure against frequent blizzards. A more in-depth, quantitative exploration of the potential influence of vulnerability to natural hazards on UDR is provided in Subsection C.

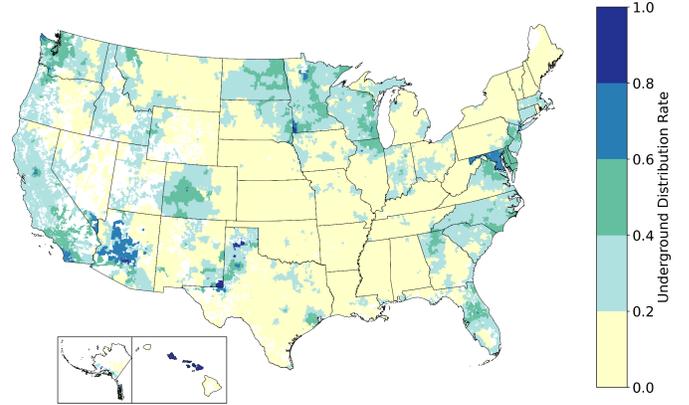

Fig. 1. Underground distribution rate at the ZCTA level in the US. White areas denote regions without ZCTAs. The progression from yellow, cyan, green, blue, and navy signifies five distinct and equal intervals of Underground Distribution Rate, ranging from low to high values.

Complementing the geographical view from Fig. 1, Fig. 2 provides a bar chart representation that associates the UDRs with the total land area and total population of ZCTAs. We found that large land areas have UDRs situated within the lower to moderate brackets of [0.0, 0.2) and [0.2, 0.4). However, as we progress to the [0.4, 0.6) range, even though the total land area diminishes sharply, the total population in these regions is more pronounced, signifying that areas with more concentrated populations are leaning towards higher UDRs. The highest bracket of [0.8, 1.0) encompasses the smallest land area with the least population, almost negligible compared with others. By comparing insights from Fig. 1 and Fig. 2, a trend that emerges is that densely populated regions, although geographically smaller, have advanced more rapidly in the adoption of underground distribution.

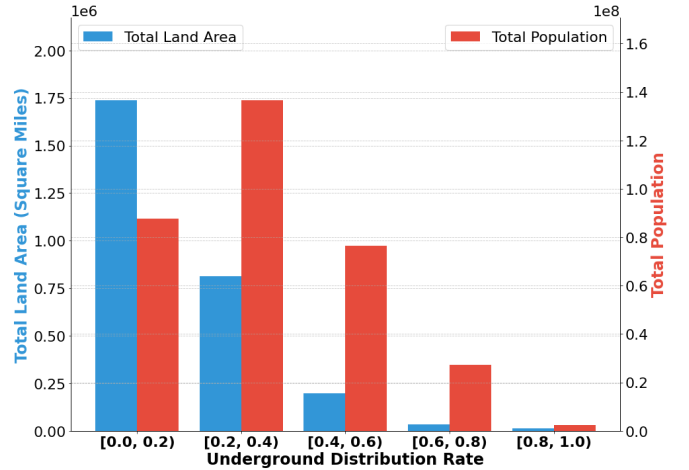

Fig. 2. Total land area and total population of ZCTAs across different underground distribution rate ranges

#### B. Underground Distribution at State Level

Building on the granular insights gained from the ZCTA level assessment, we expand our analysis to the state level to identify broader patterns and regional trends in UDRs. This perspective provides a macroscopic view, enabling a more generalized understanding of state-wise policies, investments, and preferences related to underground distribution.

The state-level relationships between total distribution mileage and UDRs across the US are displayed in Fig. 3. Washington, D.C., Maryland, and Nevada emerge as having higher UDRs, while states with extensive distribution mileage such as New York, California, and Texas display varied UDRs, suggesting that larger distribution networks do not necessarily correspond with increased or decreased undergrounding. Investigating further the interplay of these variables at the edges of the plot in Fig. 3, we found that Washington, D.C. has among the lowest distribution mileage but highest UDR, while southeastern states such as Mississippi, Louisiana, Alabama, and Tennessee have the opposite. While Rhode Island has both low distribution mileage and low UDR, there is no state at the opposite end of this spectrum with high mileage and high UGR.

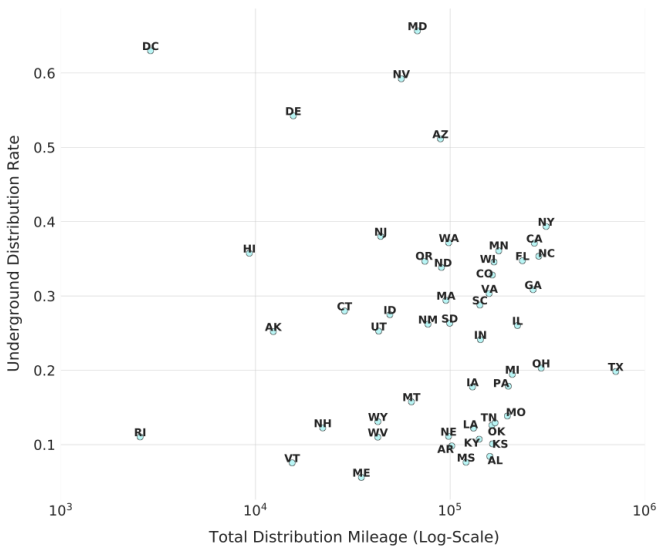

Fig. 3. Relationship between the total distribution mileage (log-scaled) and the underground distribution rate for each U.S. state. Each dot represents a state, labeled with its respective abbreviation.

## C. Factors Impacting Underground Distribution Rate

We investigated various socioeconomic and environmental metrics in relation to UDR. Fig. 4 presents a series of local polynomial regression fitting curves (LOESS), illustrating the relationship between UDR and three metrics: household mean income, RUCA, and the NRI at ZCTA level.

Fig. 4(a) displays a positive correlation between UDR and household mean income. This suggests that regions with higher mean incomes may tend to invest more in underground distribution systems, potentially due to factors like increased demand for aesthetics, land value considerations, or a preference for infrastructure resilience.

In contrast, Fig. 4(b) shows a negative correlation between the UDR and the RUCA. As we move from urban areas (codes closer to 1) to sparsely populated rural regions (codes closer to 10), the UDR tends to decrease. This trend implies that urban areas, despite their complex terrains and infrastructural challenges, may have a higher intention to implement underground distribution systems, potentially owing to land use restrictions, higher population densities, or the need to avoid overhead obstructions.

Fig. 4(c) highlights that as the NRI score (percentile ranking of NRI value among all other communities) becomes higher, there is a substantial rise in UDR. This suggests that areas with higher NRI scores, which indicates greater potential vulnerabilities to natural disasters or other hazards, may be more likely to deploy undergrounding as a hazard mitigation strategy. Underground systems, being less exposed, can generally offer increased protection against such events.

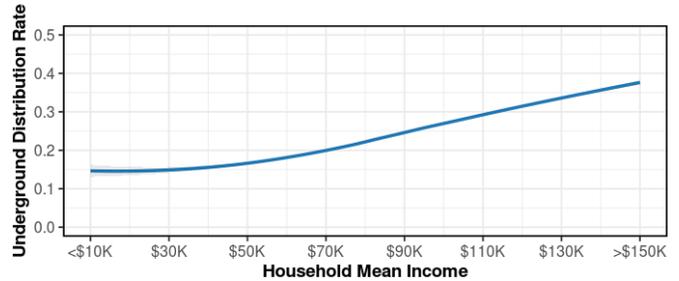

(a) Household mean income

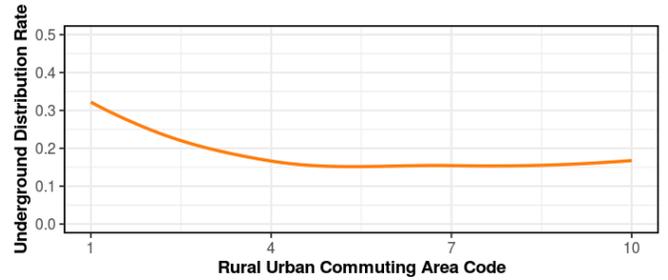

(b) Rural urban commuting area code

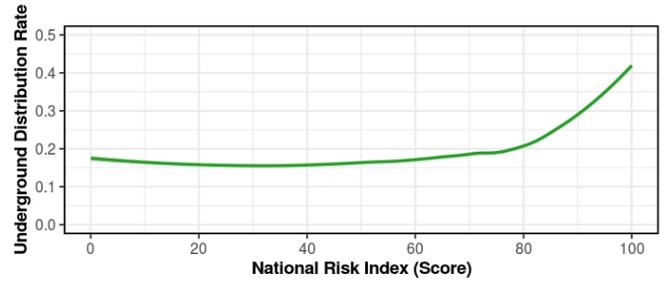

(c) National risk index (score)

Fig. 4. Relationship of Underground distribution rate with different socioeconomic and environmental metrics using LOESS. Lines represent estimates, shaded gray areas the 95% confidence interval.

Expanding on the relationships in Fig. 4, Fig. 5 provides a more comprehensive analysis of the UDR across all ZCTAs using an Ordinary Least Square (OLS) model.

The model results provide additional insights into the impact of each factor on UDP. It is noteworthy that all factors considered have statistically significant coefficient estimates ($p<0.001$). The most prominent of these factors is income, with a 1% increase in mean household income corresponding to a 0.462% increase in UDP (Fig. 4(a)). The positive estimate for population density (0.067) reaffirms the trend identified earlier in Fig. 2, suggesting that areas with denser populations have a higher proportion of underground distribution. The negative association of RUCA (-0.078) and positive estimate from NRI (0.112) also reinforce what was found in Fig.4.

Regions with higher residential rates, surprisingly, reveal a diminished UDR. Ideally, areas with high electricity tariffs should experience commensurate, if not superior, infrastructure upgrades. This trend, thus, raises questions about the equity of infrastructure distribution. While the reasons can vary—from budgetary allocations such as older infrastructure incurring higher maintenance costs to logistical challenges—these disparities may potentially magnify existing socioeconomic inequalities, and for disadvantaged communities present an additional layer of systemic neglect.

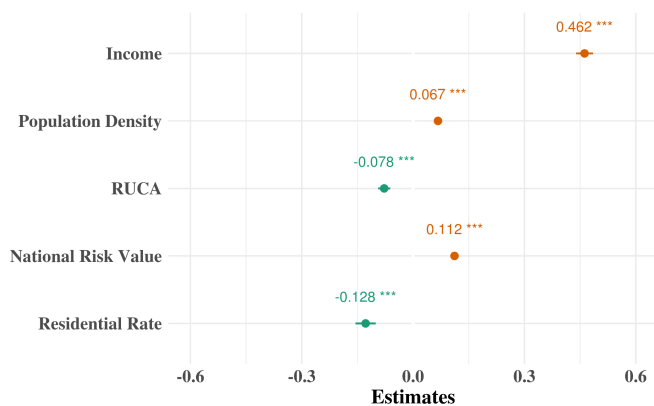

Fig. 5. Ordinary least square model predicting underground distribution rate for all ZCTAs. R-squared is 0.307. Points represent means, lines the 95% confidence interval, with statistical significance threshold denoted as * p<0.05, **p<0.01, ***p<.001.

## IV. Conclusions

Given escalating climate and energy system challenges, this study offers a granular analysis of underground power distribution across the U.S., uncovering distinct regional disparities and correlations with income, urbanization, vulnerability to natural hazards, and electricity rates. Extensions of this research could explore how undergrounding interacts with other aspects of a transitioning power grid, such as microgrids and distributed energy resources, how it is affected by various extreme weather conditions and electric shock protection, and how it will impact future reliability and energy system decarbonization.

## Acknowledgments

We would like to express our gratitude to Maryland Public Service Commission, Missouri Public Service Commission, Public Service Commission of the District of Columbia, Massachusetts Department of Public Utilities, Washington Utilities and Transportation Commission, Minnesota Public Utilities Commission, Public Utility Commission of Texas, Public Service Commission of Wisconsin, Wyoming Public Service Commission, Michigan Department of Licensing and Regulatory Affairs, Public Utilities Commission of Nevada, Illinois Commerce Commission, Kentucky Public Service Commission, California Public Utilities Commission, New Mexico Public Regulation Commission, Florida Public Service Commission, Pennsylvania Public Utility Commission, U.S. Department of Agriculture Rural Utilities Service, City of Springville, JEA, Fort Pierce Utilities Authority, City of Hermann, New Braunfels Utilities, City of Wyandotte, City of Albemarle, City of Columbus, Bay City Electric Light & Power, City of Tacoma, City of Tallahassee, City of Roseville, Knoxville Utilities Board, City of Seguin, Eugene Water and Electric Board, Brownsville Public Utilities Board, City of Forest Grove, Orlando Utilities Commission, Village of Slinger, City of Newton Falls, City of Marietta, Village of Freeport, City of Westerville, City of Rock Hill, City of Coldwater, City of Higginsville, Town of Hingham, City of Farmington, Lowell Light and Power, Holland Board of Public Works, City of Longmont, City of Gainesville, City of Independence, Mount Horeb Utilities, City of Springfield, City of Richland, Centralia City Light, City of Jackson, City of Seattle, City of Poplar Bluff, City of Salem, Stoughton Utilities, Town of Shrewsbury, City of Maryville, Kissimmee Utility Authority, City of Saint Peter, City of Logan, City of Nixa, City of Orrville, Auburn Board of Public Works, City of New Bern, City of Chaska, Town of Danvers, City of Grand Haven, Burbank Water and Power, Sterling Municipal Light Department, Alameda Municipal Power, Norwich Public Utilities, City of Cheney, City of Cody, City of Lafayette, and City of Edmond for generously providing the data essential for this study. Additionally, our sincere thanks go to Prof. Arun Majumdar, Prof. Inês Azevedo, Vermont Public Utility Commission, Maine Public Utilities Commission, American Public Power Association, Southern California Edison, City of Richland Center, and City of Marshfield for their insightful discussions that enriched the content of this paper.